\documentclass[useAMS,usenatbib]{mn2e}
\usepackage{amsmath}
\usepackage{amssymb}
\usepackage{graphicx}

\title[On the Abundance of Circumbinary Planets]{On the Abundance of Circumbinary Planets}

\author[Armstrong et. al.]{\parbox{\textwidth}{D. J. Armstrong$^1$\thanks{d.j.armstrong@warwick.ac.uk}, H. P. Osborn$^1$, D. J. A. Brown$^{1,2}$, F. Faedi$^1$, Y. G\'{o}mez Maqueo Chew$^{1}$, D. V. Martin$^3$, D. Pollacco$^1$, S. Udry$^3$}
\vspace{0.4cm}\\
\parbox{\textwidth}{$^{1}$University of Warwick, Department of Physics, Gibbet Hill Road, Coventry, CV4 7AL, UK\\
$^{2}$ARC, School of Mathematics \& Physics, Queen's University Belfast, University Road, Belfast BT7 1NN, UK\\
$^{3}$Observatoire de Gen\`eve, Universit\'e de Gen\`eve, 51 chemin des Maillettes, Sauverny 1290, Switzerland}}

\newcommand{\mytilde}{\raise.17ex\hbox{$\scriptstyle\mathtt{\sim}$}}

\begin{document}
\date{Accepted . Received}

\pagerange{\pageref{firstpage}--\pageref{lastpage}} \pubyear{2002}

\maketitle

\begin{abstract}
We present here the first observationally based determination of the rate of occurrence of circumbinary planets. This is derived from the publicly available Kepler data, using an automated search algorithm and debiasing process to produce occurrence rates implied by the seven systems already known. These rates depend critically on the planetary inclination distribution: if circumbinary planets are preferentially coplanar with their host binaries, as has been suggested, then the rate of occurrence of planets with $R_p>6R_\oplus$ orbiting with $P_p<300$\ d is $10.0 ^{+18}_{-6.5}$\% (95\% confidence limits), higher than but consistent with single star rates. If on the other hand the underlying planetary inclination distribution is isotropic, then this occurrence rate rises dramatically, to give a lower limit of 47\%. This implies that formation and subsequent dynamical evolution in circumbinary disks must either lead to largely coplanar planets, or proceed with significantly greater ease than in circumstellar disks. As a result of this investigation we also show that giant planets (${>}10R_\oplus$) are significantly less common in circumbinary orbits than their smaller siblings, and confirm that the proposed shortfall of circumbinary planets orbiting the shorter period binaries in the Kepler sample is a real effect.
\end{abstract}


\section{Introduction}
\label{sectIntro}
In recent years the incredibly precise Kepler data has produced a wide range of important discoveries. Among these is the array of planets now known orbiting binary stars, proving not only that circumbinary planets can exist stably in such locations, but that they are not rare. At the time of writing seven systems with transiting circumbinary planets are known, being Kepler-16b \citep{Doyle:2011ev}, Kepler-34b and -35b \citep{Welsh:2012kl}, Kepler-38b \citep{Orosz:2012ip}, Kepler-47b and c \citep{Orosz:2012ku}, Kepler-64b/PH1 \citep{Kostov:2012vb,Schwamb:2012ts} and the recently published Kepler-413b \citep{Kostov:2014bx}. Although there are significant obstacles to routine detection in the form of large transit timing and duration variations \citep{Armstrong:2013kg}, the relatively small sample of Kepler eclipsing binaries has produced a sizeable number. Several questions remain: how abundant are these planets? How does the central binary affect their formation \citep[e.g.][]{Pelupessy:2013gl}? What evolutionary processes dominate in such an environment \citep[e.g.][]{Pierens:2008fq,Pierens:2013ee}?

Some theoretical work has been done, showing that planet formation in a circumbinary disk could be hindered by raised planetessimal velocities \citep{Meschiari:2012er} over an area several AU in size, including the present orbits of the known planets. This implies that circumbinary planet formation may well proceed on wider orbits, with planets later migrating to their current positions \citep{Kley:2014vv}. The exact extent of the formation suppressing area is as yet unknown, and it has been proposed that planet formation in circumbinary disks may be helped by zones of lower velocity \citep{GMartin:2013be,Rafikov:2012wl}. How easily such planets form, and the evolutionary route they follow, represents an excellent constraint on planet formation in general.

The Kepler sample of circumbinaries (CBs) has grown to a point where it can begin to tell us about these planets in general. Here we use it to extract what information we can on the rate of occurrence of CBs, as well as their distribution of inclinations. These are important indicators of the history of CB systems, showing whether formation proceeds easily of with difficulty, and whether scattering plays a key role in any subsequent evolution. We focus here on detached binary systems, and on planets with periods within 300\ d. For reasons of completeness we only utilise planets showing consecutive transits, i.e. those which produce a transit on each orbit. There are expected to be many `sparsely' transiting CBs that only occasionally transit \citep{Martin:2014ug}, and which can be expected to provide more information in future. We use occurrence rate here to mean the number of binaries with one (or more) planets as a fraction of the total binary number, leaving the question of multiple planets per binary to future work.

\section{Data Processing}
\subsection{Data Source}
\label{SectDataSource}
Targets were selected from the Kepler Eclipsing Binary Catalogue (KEBC) \citep{Slawson:2011fg}. The version of the catalogue as found online\footnote{http://keplerebs.villanova.edu} on 18th September 2013 was used, yielding 2610 objects. Of these we restricted ourselves to systems with morphology parameter ${<}0.7$ (i.e. detached and semi-detached binaries, see \citet{Matijevic:2012di} for detail), as the history of planets in highly evolved over-contact binaries is likely to be significantly different from those in other systems, and these binaries present different challenges to systematic planetary detection. Our initial sample then comprised of 1735 binaries. Light curves from Quarters 1 through 16 were downloaded, comprising a baseline of approximately 4 years for most objects.

The KEBC provides us with period information for the binary sample. In addition data is available on the locations and widths of the primary and secondary binary eclipses. With these, eccentricity parameters were calculated. Where insufficient polyfit information was available (generally due to non-detection of the secondary eclipse) we set the binary eccentricity to zero. Temperature information on the sample was obtained from \citet{Armstrong:2013cp}. We generated main-sequence calibrated stellar radii and masses from these, using the calibration of \citet{Torres:2010eoa} with surface gravity 4.5 and solar metallicity. The lower mass limit for this process was 0.6$M_\odot$ - below this the calibrations of \citet{Boyajian:2012eu} were used. Between 0.6 and 0.8$M_\odot$, in the valid range of both calibrations, we interpolated between them to ensure no discontinuity.

\subsection{Data Detrending}
We elected to detrend the light curves from instrumental and systematic effects using covariance basis vectors. These were used over the research data conditioning (PDC) detrended data available as the PDC data is not robust against long-duration events, as warned in \citet{KeplerDatProcess}, and the transits of CB planets may in theory last for half the orbital period of the binary. While it would be ideal to individually tune the detrending of all light curves, the sample size made this impractical. Detrending was enacted using the PyKE code \citep{Still:2012tv}. At this stage, data with a non-zero SAP\_QUALITY flag was cut (see \citet{KeplerArchiveManual} for full list of exclusions). Once detrended, quarter data was stitched together through dividing by the median flux value of each quarter, forming single light curves for each binary.

\subsection{EB Signal Removal}
\label{sectebsigremoval}
At this stage the signal of the known eclipsing binary must be removed, without affecting any potentially planetary signals. We do this using a modified whitening procedure, whereby the light curve is phase-folded at the binary period. The phased curve is then binned into equal width bins, and the median of each determined. Bins with significantly higher than average variance are then further subdivided into higher resolution bins, mitigating the effect of sharp variations. The median of each bin is then subtracted from each point in the bin. As any points showing tertiary signals will be distributed through the phase curve, taking the median excludes them from the process. This method has the disadvantage of occasionally leaving residual binary signals around regions of sharp variation, particularly ingress or egress points for detached binary eclipses. To lessen this effect, each curve was checked by eye for strictly periodic binary remnants, and any found manually removed.

\section{Search Algorithm}

In order to avoid the subjective nature of searching the light curves by eye, an automated search algorithm was used on the dataset. This composed two stages: a search for significant individual transit-like events, followed by a periodicity test. 

\subsection{Individual Event Search}
\label{sectindieventsearch}
To test for individual transit events, a box was passed across the light curve at a 0.1\ d resolution (i.e. 4-5 data points). Gaps in the data are often surrounded by systematic noise; as such, 0.5\ d regions around gaps (defined by a greater than 0.5\ d space between two adjacent points) were ignored. Points falling at known binary eclipse times were also screened (only for well detached binaries with morphology ${<}0.2$, such that the eclipses could not take up a large proportion of the light curve). At each step, a 3\ d window centred on the current box was taken. Three days was chosen to give a significantly long baseline, while still being short enough to track variability. Periodic noise or stellar variability with timescale less than the baseline fitting region will obscure planetary signals. A third order polynomial was fit to this region, excluding the central 0.1\ d box. Gaps were not fit across, due to discontinuities in the data often marked by a significant gap. This fit was repeated for 20 iterations, with points $>5\sigma$ from the best fit excluded each time. The offset of the central box from the best fit baseline, relative to the noise of the 3d region around the best fit baseline, was then taken and stored. After the whole light curve is tested, any times with offset significances $>3\sigma$ of the whole set of significances are passed on to the periodicity test.

\subsection{Periodicity}
Due to the large transit timing variations (TTVs) on the order of several days in circumbinary planet signals, events cannot be held to be strictly periodic and the usual methods for forming periodograms cannot be used. We test for periodicity by phase-folding the central times of each detected event at a series of trial periods, using the same method as \citet{Armstrong:2012ie} and similarly to a box-least-squares search \citep{Kovacs:2002ho}. At each period we test for groupings of event times, within a box width defined by the maximum possible TTV for the specific binary. This maximum is derived in \citet{Armstrong:2013kg}; we consider only the geometric contribution to the TTVs (i.e. from the motion of the stars, and ignoring planetary precession), taking parameters from the KEBC, as well as eccentricity parameters as described in Section \ref{SectDataSource} for each binary. We assume stellar mass ratios in their formulae to give the largest, and therefore most robust, upper limit on the TTVs. Each event time is weighted by the significance of its detection, and our test periods range from 320\ d down to either 2\ d or 2.5 $P_\textrm{bin}$, whichever is longer. The 320\ d limit is used so as to avoid a hard periodogram limit at the same point as our limit for statistical purposes. It is well below the full duration of the Kepler data, which allows for at least 4 orbits of a 300\ d period planet. Two days is where individual transits may become hard to detect in the long cadence (30 min resolution) Kepler data, and 2.5 $P_\textrm{bin}$ is set such as to be well within the inner stability limit given by \citet{Holman:1999cu}, which is typically 4-5 $P_\textrm{bin}$ for circular binaries. The total significance within the box is then saved, forming a periodogram over the whole tested range. As the maximum TTV (and hence box width and so number of data points contained within the box) grows for smaller planet to binary period ratios, a preference for shorter periods is introduced; we remove this by applying a weighting of the inverse of the box width.

\subsection{Output Statistic and Detections}
\begin{figure}
\resizebox{\hsize}{!}{\includegraphics{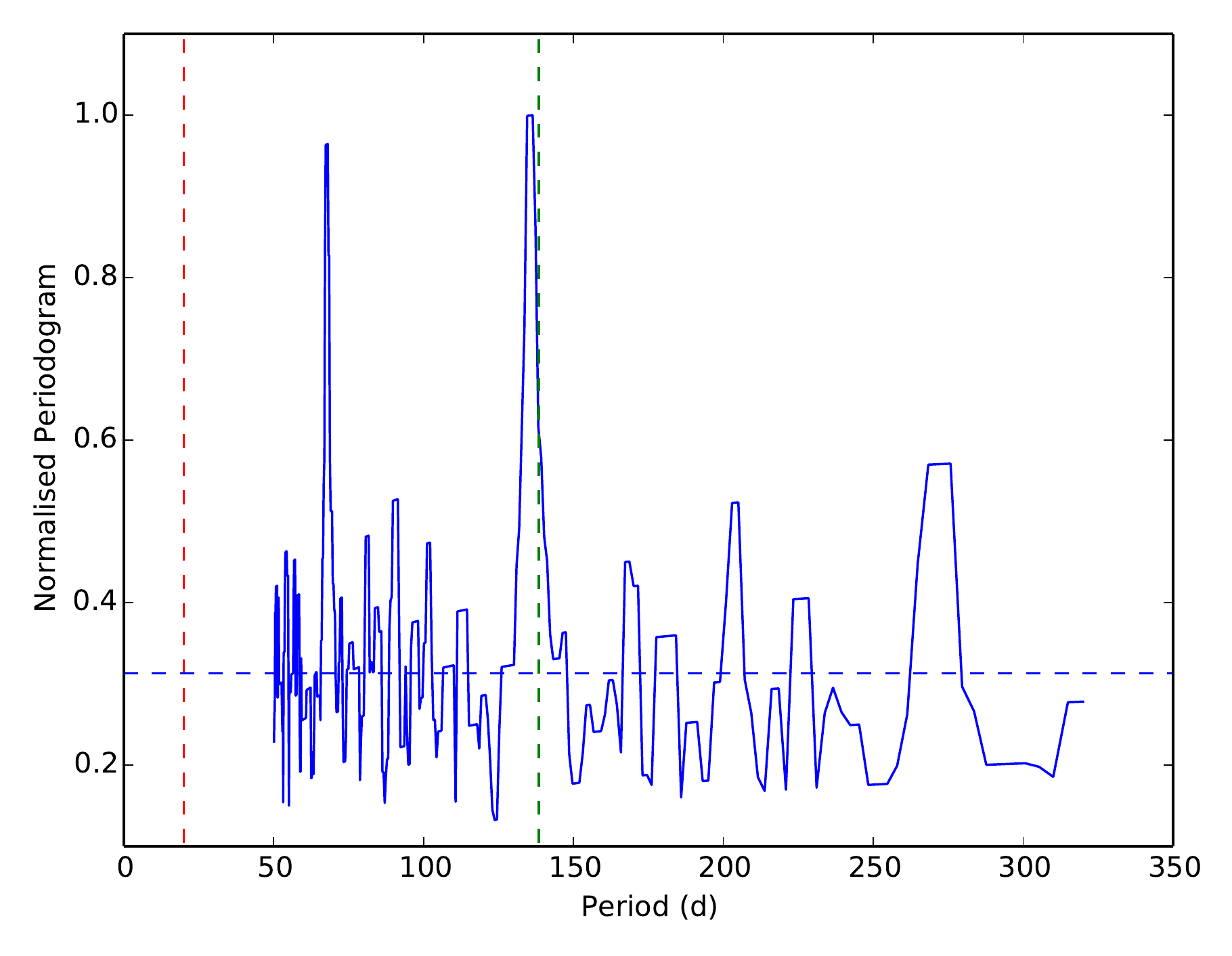}}
\caption{The detected periodogram for the known planet PH1. Note the strong P/2 and 2P harmonics. The correct period is shown as a green dashed line (near 140\ d, and the binary period as red, near 20\ d. The correct period shown is the published period - this is a few days larger than the azimuthal period, which represents the mean transit interval and is what is detected by the search algorithm. The median value of the periodogram is shown as a horizontal dashed line, and with the peaks led to a detected significance of 2.7}
\label{figperph1}
\end{figure}

Due again to the large TTVs, and the possibility of multiple transits appearing on a single planetary orbit \citep{Liu:2014ta}, we must adopt a more unusual method for forming an output statistic. In particular these issues combined with this search algorithm lead to a tendency to detecting harmonics - here the maximum peak of a typical planet detection periodogram is often a harmonic of the true period. This is due to the maximum TTV region tested by the algorithm at each period. If any planetary transits do not completely fill this region, it is possible for harmonics (which have a different maximum TTV, but one which may still cover all the known transits) to give strong periodogram peaks. This effect is not so prevalent for noise, and so the presence of strong harmonics can be used to advantage -- when finding an output statistic we take account of those at $P_p$/2, $P_p$/3, 2$P_p$ and 3$P_p$, where $P_p$ is the tested planet period. The mean value of the maximum peak and these harmonic peaks was taken, each divided by the median value of the periodogram to take account of the different levels of noise between objects, producing the detected significance. Using additional harmonics to form the average was found to not significantly improve the results. A periodogram for one of the known planets, PH1, is shown in Figure \ref{figperph1}. Note that in this case the $3P_p$ and $P_p/3$ harmonics are out of the tested range and so were not used.

\begin{figure}
\resizebox{\hsize}{!}{\includegraphics{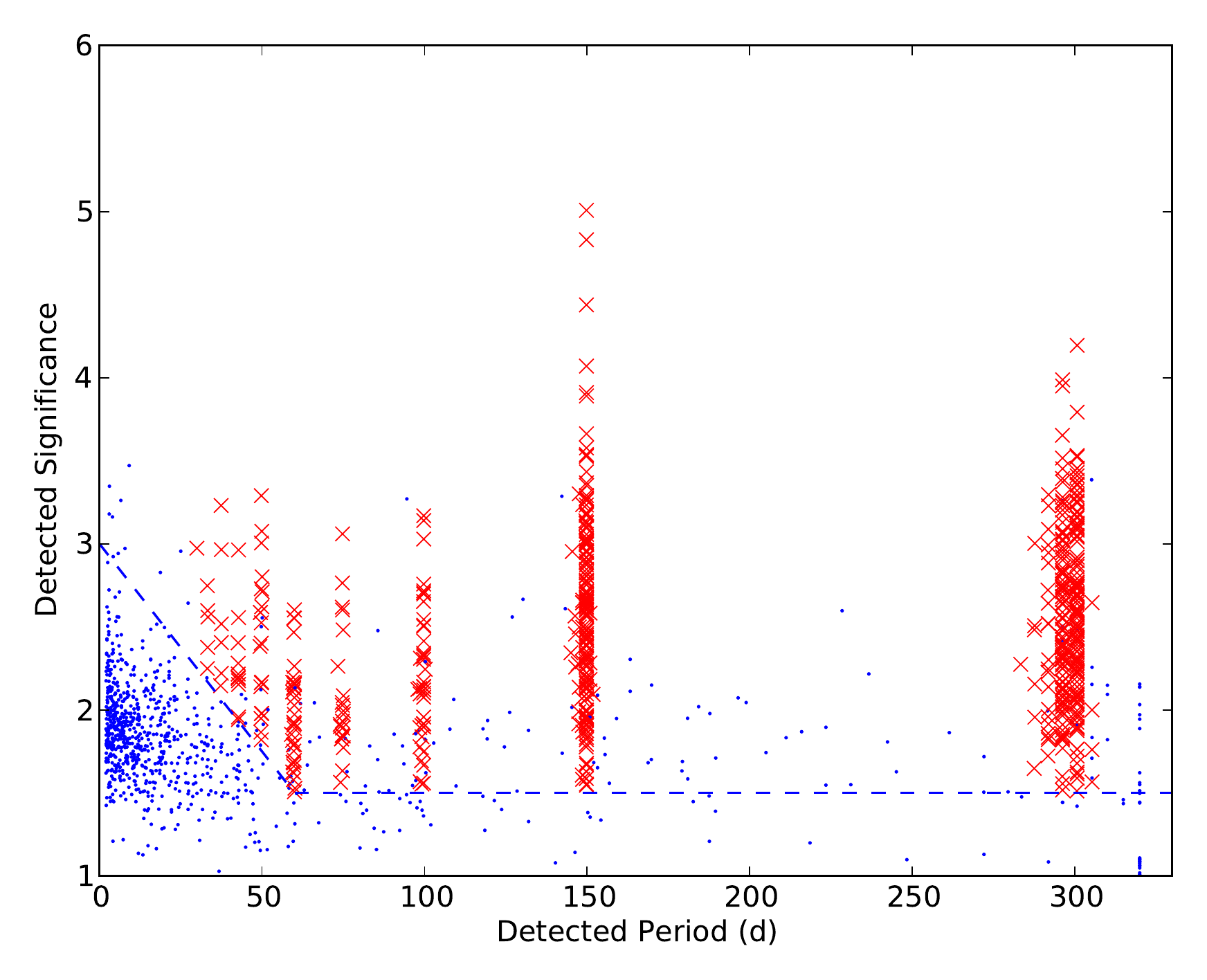}}
\caption{Detected significance of the whole binary sample with injected planetary signals, for $P_p=300$d, $R_p=10R_\oplus$. Red crosses represent successful detections at period or harmonic, while blue dots represent the detected significance of the highest peak in the periodogram in systems where the injection was not recovered. Dashed line represents the significance threshold used. Some blue dots fall above this line due to chance periodicity in the noise (be it astrophysical or instrumental) of those system's lightcurves.}
\label{figinjectionsignificance}
\end{figure}

A minimum significance threshold was set using the significances of recovered test transit injections (see Section \ref{SectTInjections}) and is shown in Figure \ref{figinjectionsignificance}. Note the large number of unrecovered injections at periods below 60\ d -- due to the increased box size at short periods (caused by increasing maximum TTV), both the number and significance of false detections is increased in this region. As shorter period real signals have their significance increased for the same reason, as well as having more transit events, we set a minimum significance threshold which rises at low periods to exclude this additional noise. This led to 308 out of 1735 systems showing signals with significance over the threshold, which were in each case examined transit by transit by eye. Further to this, every periodogram and light curve was checked by two independent researchers. Both flagged all of the currently known planets clearly, with the exception of the Kepler-47 system, which was only weakly detected due to both small transits and stellar noise. We also detected two strong candidate planetary systems within our period threshold of 300\ d as well as various other signals both potentially planetary and not. These are described in Section \ref{SectResultsCandidates}.

\section{Debiasing}
\label{SectDebiasing}
\subsection{Transit Injections}
\label{SectTInjections}
To test the efficacy of our search algorithm and determine what the true sample of searched light curves was, we created simulated planetary signals and attempted to recover them. Transit times were found using an N-body integration code, in which binaries with the mass ratios, stellar radii, periods and eccentricities of the KEBC systems (see Section \ref{SectDataSource}) were created, then planets placed around them. This allowed for the inclusion of the various non-Keplerian orbital effect found in circumbinary systems \citep[e.g.][]{Leung:2013fm,Li:2013do}. The planets were put in orbits coplanar with the binaries on periods of 10.2 $P_\textrm{bin}$ and 300\ d. These were made slightly eccentric (e=0.05, with a uniform distribution of argument of periapsis between 0 and $2\pi$) so that the additional TTVs which eccentricity may bring were not excluded. Systems which were unstable (typically very long period binaries where 300d proved to be within the inner stability limit) were dropped, providing an implicit stability check on our sample. Exact resonances with the binary were avoided, due to the possibility of localised stability effects \citep{Doolin:2011ib}. Times and durations of transits were extracted. 

These transits were injected into the light curves of each binary using U-shaped transits of the recorded duration, centred on the transit times. Only transits of the primary star were used, as these dominate the detectability of a planet. Transit depths were set using stellar radii derived as described in Section \ref{SectDataSource}, for planets of radius $4R_\oplus$, $6R_\oplus$, $8R_\oplus$ and $10R_\oplus$. Dilution from the secondary star was included, along with quarter-by-quarter contamination ascertained from the Kepler data archive \citep[][typically a few percent]{KeplerArchiveManual}. Note that contamination by unknown tertiary stars in the system is not included, as no information is available as to the extent or magnitude of this. Although approx. 20\% of the KEBC binaries are thought to have stellar tertiary companions \citep{Rappaport:2013de} the amount that these will dilute transits of the primary binary star is unknown. Each planetary radius and period combination was searched separately. The detected output statistics for a typical injection group are shown in Figure \ref{figinjectionsignificance}, and led to the threshold shown. Injections where the maximum periodogram peak was not at a harmonic of the injected signal, or where the detected significance was below the threshold, are shown as blue dots and represent the background noise distribution. Detections were allowed for any harmonic down to $P_p/10$. Signals on shorter periods show more events and as such generally have higher detection significances. The effect was particularly strong for signals under 60\ d, which is reflected in our threshold.

Similarly, light curves containing known planets or candidates were subjected to testing, after removal of the already seen transits. This allowed us to probe the sensitivity of the search algorithm in these systems. Table \ref{tabplanetsample} shows the number of planets or candidates contained within each period or radius bin for which these test injections were recovered successfully. Note that some planets which would be expected to appear in bins do not because we did not successfully recover transit injections at those minimum radii in these systems. An example is Kepler-34b in the $4-10R_\oplus$ group, where a 4$R_\oplus$ transit injection was not recovered and so the system is not included, as in the Kepler-34 system a planet could not be detected over the whole bin range. On the other hand Kepler-16b is included in the $4-10$ bin, as a test $4R_\oplus$ planet was successfully recovered for this system and the real planet radius lies within the bin.

\begin{table}
\caption{Planets and candidates within bins for which test transit injections were successful. \textrm{Note that the 8-10 bin is equivalent to $>8$ here.}}
\label{tabplanetsample}

\begin{tabular}{@{}lllr@{}}
\hline
Period & Radius & Number & Included Planets \\
         &  ($R_\oplus$)   &   &     \\
\hline
\hline
10.2 $P_\textrm{bin}$ & $>10$ &  0 & \\
 & $8-10$ &  2 & K-16b, K-35b\\
 & $6-8$ &  1 & PH1\\
  & $4-6$ &  2 &  K-38b, KIC6504534\\ 
 & $6-10$ &  3 & K-16b, K-35b, PH1\\
  & $4-10$ &  3 & K-16b, K-38b, KIC6504534\\
300\ d   & $>10$ &  0 & \\
    & $8-10$ &  3 & K-16b, K-34b, K-35b\\
 & $6-8$ &  1 & PH1\\
  & $4-6$ &  2 &  K-38b, KIC6504534\\ 
     & $6-10$ &  3 & K-16b, K-34b, PH1\\
      & $4-10$ &  3 & K-16b, K-38b, KIC6504534\\
\hline

\end{tabular}

\end{table}

\subsection{Test Results}
The number of systems where recovery was successful according to the stated threshold is shown in Table \ref{tabsamplesize}, split into each radius and period group. This includes systems where the detected period was a harmonic of the injected period. Our recovery rate varied between \mytilde10\% of the total stable sample for the most difficult 300d, 4$R_\oplus$ case, and \mytilde55\% for the 10.2$P_\textrm{bin}$, 10$R_\oplus$ case. To check these surprisingly low sample sizes, a subsample of the failed systems were examined to determine the cause of the recovery failure. In \mytilde50\% of $R_p = 10R_\oplus$ cases this was light curve noise or stellar activity dominating the transit signal depth. A further sixth of the failed cases were due to remnants of binary eclipses, with another sixth due to short light curves (generally under 1 year) which were not long enough to show multiple transits. The remainder were due to transits falling in gaps in the light curve, with a few percent finding the correct injected period but at too low significance. We note that dilution by the secondary star in general had a large effect on the transit depths, resulting in transits significantly shallower than would be expected for e.g. $10R_\oplus$ planets around single stars. In the following, we assume that a system which tested successfully at a given planet period and radius would also be successful at any shorter period or larger radius, as both of these changes make detection easier. Note that this method allows us to use the specific sample of binaries in which we could detect planets, so that we are finding the implied occurrence rates truly given by this sample. As such while the recovery percentages give a good idea of completion rates, it is the specific binaries which make up each sample, and moreover the parameter space of each within which planets would be observable as found in Section \ref{sectpopsynth}, that are the most important outcome of this debiasing process.

\begin{table}
\caption{N systems with successful transit injection recovery, out of 1735 total.}
\label{tabsamplesize}

\begin{tabular}{@{}llr@{}}
\hline
Period & Radius & N$_\textrm{recovered}$ \\
         &  ($R_\oplus$)   &       \\
\hline
\hline
10.2 $P_\textrm{bin}$ & 10 &  857 \\
 & 8 &  757\\
 & 6 &  597\\
  & 4 &  322\\
300\ d   & 10 &  581\\
    & 8 &  490\\
     & 6 &  328\\
      & 4 &  143\\
\hline

\end{tabular}

\end{table}

\section{Population Synthesis}
\label{sectpopsynth}
\subsection{Overview}
Converting the sample size and number of observed planets that we have into useable statistics requires some synthesis of circumbinary planet populations.This proceeds as a separate step in the method to the debiasing of Section \ref{SectDebiasing}, with the only input being the specific sample of binaries for each parameter bin as well as the corresponding known planet number. The aim of this Monte Carlo based population synthesis is to find what occurrence rates with what probability are consistent with this debiased sample and known number of consecutively transiting planets. These then form posterior probability density functions for the occurrence rate. They will vary with the underlying planet distributions. This is because, while the binary sample and planet count are fixed, many unobserved planets may exist, especially in the higher inclination regions where much (if not all) of the planet's possible orbital parameter space will not produce consecutive, or indeed any, transits. We do not attempt to perform a completeness adjustment through analytically finding what region of parameter space is covered by consecutive transits and adjusting by that. Such an adjustment is performed implicitly by this method, which simultaneously finds errors on the derived values through the probability functions output.

We simulated planets orbiting our sample binaries using a Keplerian approach. While this ignores the more complex dynamics of circumbinary planetary systems, including rapid precession, period and eccentricity oscillations \citep[e.g.][]{Farago:2010ev,Doolin:2011ib,Leung:2013fm}, the approximation must only hold for the time baseline of 4 years which we use. Furthermore, the produced planet count would only be sensitive to systematic offsets caused by these effects, which we expect to be small, rather than orbital element variation which would be taken account of when distributing the orbital elements. This approximation allows us to rapidly sample many possible combinations of orbital elements, something that would be both time and computationally expensive using a full N-body simulator. Under this system binaries are placed into their known Keplerian orbits, and planets then simulated orbiting the system barycentre. We restricted ourselves to circular planets. This should have little to no effect on the results, as while some slightly inclined planets on circular orbits would stop transiting consecutively if made eccentric, a similar number which did not previously transit consecutively would begin to (given a uniform distribution of argument of periapse.)

\subsection{Planet Distributions}
\label{sectpopsynthplanetdists}
\begin{figure}
\resizebox{\hsize}{!}{\includegraphics{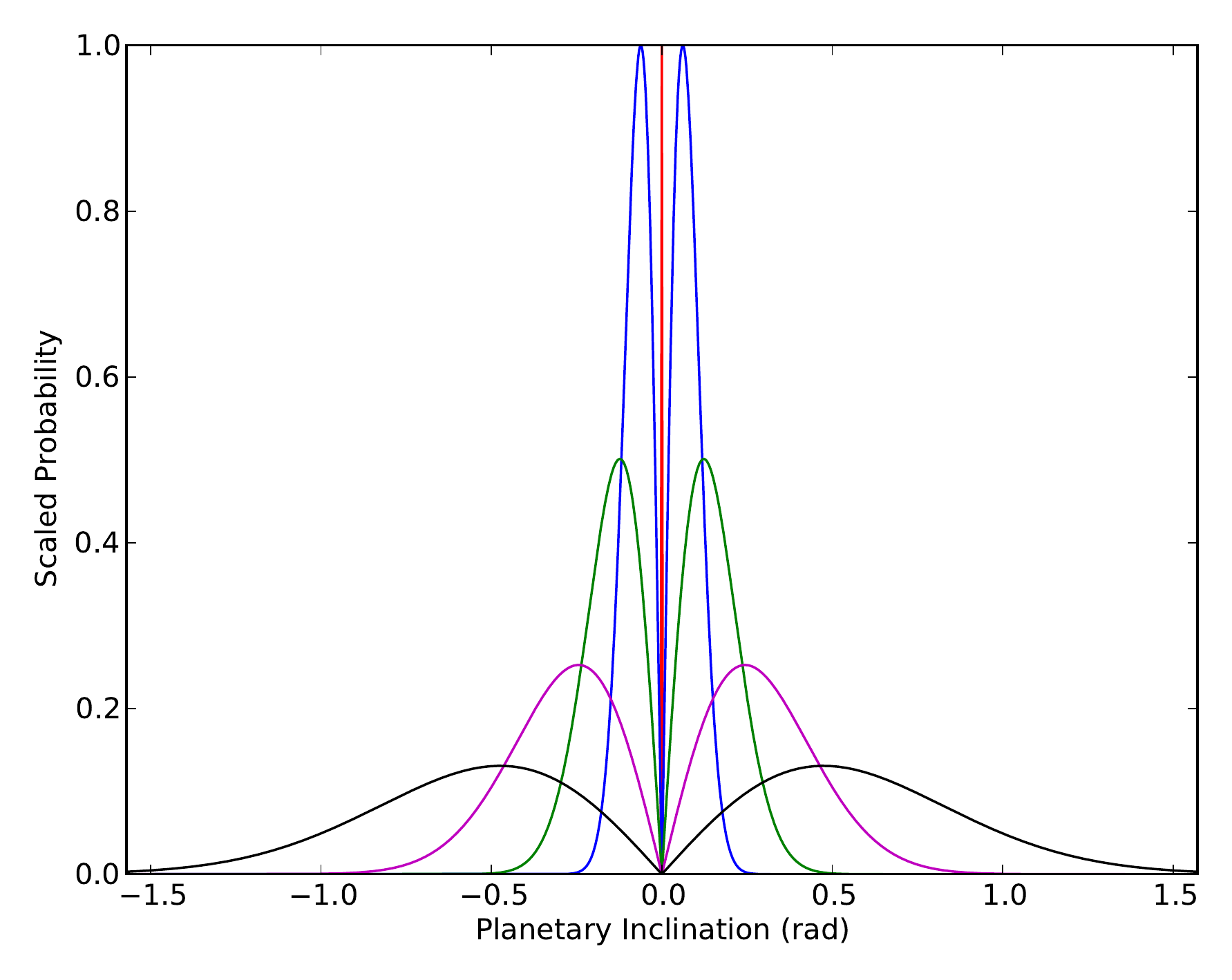}}
\caption{Probability density functions for the synthesised planet population inclinations. Distributions are (from centre out) Coplanar, then Gaussian 5, 10, 20 and 40 $^\circ$ (See text). These are normalised such that the Gaussian $5^\circ$ distribution peaks at unity.}
\label{figincdists}
\end{figure}

For planets the crucial distributions are those of inclination and period. There are theoretical indications that planetary inclinations should be preferentially coplanar with the binary \citep{Foucart:2013gk}. The actual distribution is largely unknown, with influences possible from protoplanetary disk alignment, planet-planet scattering \citep[e.g.][]{Chatterjee:2008gd} and other sources of orbital evolution \citep[e.g.][]{Kley:2014vv}. If all circumbinary planets were near perfectly aligned with their binary orbital planes, our detected numbers would represent a significantly different underlying abundance than if the planets were uniformly distributed. As such we test a variety of inclination distributions, and present occurrence rates as a function of these. All inclinations are measured relative to the binary plane. We trial gaussian distributions with means of zero and standard deviations ranging from 5 to 40 degrees. These are simple functions which can easily be made `more misaligned', and so without better knowledge of the true distribution represent a good test case. Each of these is convolved with the standard isotropic uniform in cos i distribution (i.e. convolved with sin i at the probability distribution stage). This is done to avoid the bias towards values near zero which would result from using the gaussian distributions directly. We also test an isotropic distribution, as well as a fully coplanar distribution. The injected distributions are shown in Figure \ref{figincdists}.

In terms of planetary period, the underlying distribution is again poorly known. Using the justification that far from the central binary planet formation and evolution can be expected to proceed as if the host was a single star, we use the distribution of periods found from the Kepler objects of interest (cut off above 300\ d, and corrected for the reduced probability of long period planets transiting), and without further knowledge assume this distribution holds down to the inner stability limit \citep{Holman:1999cu,Dvorak:1989wm} of each binary, below which planets are taken to be ejected or absorbed by a host star. There have been indications \citep{Welsh:2013wo} of a potential `pileup' of planets close to this stability limit - for example through the halting of inward migration at the disk boundary. As such, we also trialled a distribution whereby 50\% of the planets located within the inner stability limit are `recovered' and placed randomly between 1.1 and 1.4 multiples of that limit. Our results proved to be generally insensitive to this, and so final results are presented without this pile-up.

\subsection{Binary Distributions}
Many of the necessary binary parameters are already known and were used as described in Section \ref{SectDataSource}. Binary inclinations were drawn uniformly across the range within which they would still at least partially eclipse. It is critical to include the binary inclination variation, as for preferentially coplanar planets on much larger semimajor axes a change of a few degrees can have significant consequences for observability.

\subsection{Output}
In each iteration, a proportion of binaries are assigned a planet randomly based on a tested occurrence rate. At this point the relevant binary and planet parameters are drawn, and planets then checked for observability. We limit ourselves to planets transiting consecutively, i.e. on every orbit, as while Section \ref{SectDebiasing} makes our sample complete for consecutive transits, the effect of occasional missed transits is difficult to quantify. This constraint requires a high degree of alignment with the binary plane, to a degree commensurate with the stellar radii. We then obtain a total number of observable, consecutively transiting planets, for a given occurrence rate and set of parameter draws. Iterated 10,000 times, we gain a distribution of observable planets for the tested occurrence rate, and then repeat for a range of occurrence rates. At this point we have a probability distribution of the expected number of consecutively transiting planets for each tested occurrence rate. These can then be inverted - taking our known number of planets (see Table \ref{tabplanetsample}) we can see how many times this number was found for each rate, thereby producing a distribution of occurrence rates for a given planet count. Assuming a uniformly distributed prior on the occurrence rate (as is appropriate given the current lack of observational evidence), these can be normalised to form probability density functions.

\section{Results}

\subsection{Detected Tertiary Signals}
\label{SectResultsCandidates}

\begin{table}

\caption{Candidate Planets}
\label{tabcandplanets}
\begin{tabular}{@{}lllr@{}}
\hline
Kepler ID & P$_\textrm{bin}$ & e$_\textrm{bin}$ & P$_\textrm{candidate}$ \\
                 &  (d)        &   &   (d)  \\
\hline
\hline
5473556 & 11.26 & 0.15 & 550 or 1110 \\
6504534 & 28.16 & 0.094 & \mytilde 170\\
9632895 & 27.32& 0.093 & \mytilde 240\\
\hline
\end{tabular}
\end{table}

\begin{table}

\caption{Other Signals}
\label{tabothersignals}
\begin{tabular}{@{}llll@{}}
\hline
Kepler ID  & Comment\\
\hline
\hline
6144827  &   Additional eclipsing binary (EB) signal at 1.94d\\
7871200  &  Highly eccentric ($e \simeq 0.7$) additional EB \\
 & signal at 38.02d\\
8113154  &  Broad 5d long faint regions on 40d period\\
10223618  &  For several consecutive quarters binary\\
&  secondary eclipses gain an additional\\
&  1\% dip just after eclipse  \\
\hline
\end{tabular}

\end{table}

We detected three strong candidate planetary systems, two of which lay within our period limit of 300\ d. These are in addition to the currently known planets and Kepler-413b \citep{Kostov:2014bx}, all of which were strongly detected, excepting Kepler-47. We do not analyse or attempt to confirm our candidate systems, as this is beyond the scope of this paper. We can however give approximate periods of the planets these systems would represent if they prove real. Table \ref{tabcandplanets} gives our input and derived parameters for these candidate systems. 

The shortest period object, KIC6504534, shows three clear transits, with several gaps where others would be expected to fall. These transits imply a \mytilde170.3d planet, showing transit timing variations of at least 0.1\ d, as well as transit duration variations of a similar magnitude. Using our calibrated stellar radii (see Section \ref{SectDataSource}), the transit depth of \mytilde0.2\% would represent a planet radius of 4.3 $R_\oplus$. While this planet is not confirmed, the presence of clear transits with strong timing and duration variations supports the hypothesis that it represents a real signal. The candidate period shown by its transit signals also corresponds to \mytilde 6$P_\textrm{bin}$, similar to the currently known planets and outside the inner stability limit for this system. Given these considerations, we include KIC6504534 in the planet count when calculating rates of occurrence. 

KIC 5473556 was mentioned in \citet{Welsh:2012kl} as showing a single transit. There are now two, implying a period of 550 or 1100\ d (due to a gap in the light curve where a transit could have been missed). This candidate does not have enough transits to show TTVs, leaving the possibility of a background blend open. Our remaining candidate, KIC9632895, shows three transits, implying an \mytilde 240\ d period with TTVs of magnitude over 1d. There are however light curve regions where consecutive transits should lie, implying that this candidate is on a slightly misaligned orbit. As such it is not within our consecutive transit threshold, and is not used to compute planet occurrences.

We also detected several eclipses too deep to be planets. Many of these are already known multiple star systems, and we will not list them here. However a few other and as far as we are aware unknown signals were also found, and these merit noting. They are summarised in Table \ref{tabothersignals}. We make no comment on the possible nature of these objects -- there is a significant chance that some of them represent blended background source for example, but others may be triple star systems or simply stellar activity.

\subsection{Occurrence Rates}

\begin{table*}

\caption{Percentage rates of occurrence for planets within 10.2$P_\textrm{bin}$. Values are maximum likelihoods, with the rates of occurrence corresponding to 50 and 95\% confidence intervals shown as super and subscripts.}
\label{taboccrates102}
\begin{tabular}{@{}llllllr@{}}
\hline
R$_\textrm{planet}$ & \multicolumn{3}{|l|}{Planetary Inclination Distribution} \\
(R$_\oplus$) &  Coplanar  &    Gauss $\sigma=5^\circ$  & Gauss $\sigma=10^\circ$  & Gauss $\sigma=20^\circ$  & Gauss $\sigma=40^\circ$  & Isotropic \\
\hline
\hline
\vspace{0.5em}
$>10$ &  $0\>^{0.67}$  $^{2.8}$  &     $0^{1.1}$  $^{4.8}$  &  $0\>^{1.9}$  $^{8.0}$  & $0\>^{3.7}$  $^{15.9}$  & $0\>^{9.0}$  $^{39}$  & $0\>^{26}$  $^{84}$  \\

\vspace{0.5em}
$8-10$ &  $2.0\>^{4.2}_{1.8}$  $^{7.6}_{0.57}$  &  $3.7\>^{7.6}_{3.3}$  $^{13.9}_{1.1}$    & $7.1\>^{12.9}_{5.7}$  $^{24}_{2.0}$   &$15.0\>^{26}_{11.7}$  $^{48}_{4.1}$    & $33\>^{60}_{28}$  $^{92}_{10.1}$  & $100\>_{70}$  $_{27}$ \\

\vspace{0.5em}

$6-8$ &  $1.5\>^{3.6}_{1.3}$  $^{7.5}_{0.25}$  &  $2.5\>^{6.7}_{2.4}$  $^{13.9}_{0.59}$  & $5.0\>^{11.4}_{4.1}$  $^{24}_{1.0}$ & $9.0\>^{8.4}_{23}$  $^{48}_{2.1}$ & $23\>^{53}_{19.9}$  $^{90}_{5.1}$ & $75\>^{79}_{39}$  $^{97}_{10.9}$ \\

\vspace{0.5em}

$4-6$ &  $5.0\>^{9.4}_{4.2}$  $^{17.3}_{1.4}$  &  $9.5\>^{18.0}_{7.9}$  $^{33}_{2.8}$  & $15\>^{31}_{13.7}$  $^{56}_{4.9}$ & $31\>^{60}_{27}$  $^{91}_{10.0}$ & $86\>^{100}_{66}$  $^{100}_{24}$ & $100\>_{76}$  $_{33}$ \\

\vspace{0.5em}
$6-10$ & $4.2\>^{6.8}_{3.4}$  $^{11.7}_{1.4}$    &  $7.5\>^{12.7}_{6.3}$  $^{22}_{2.7}$   & $13\>^{22}_{10.7}$  $^{37}_{4.5}$   & $25\>^{44}_{22}$  $^{73}_{9.4}$   & $68\>^{80}_{46}$  $^{97}_{21}$   & $100\>_{79}$  $_{40}$  \\

\vspace{0.5em}
$4-10$ & $7.1\>^{12.2}_{6.1}$  $^{21}_{2.6}$   &  $13.5\>^{24}_{11.7}$  $^{40}_{5.0}$   & $25\>^{40}_{20}$  $^{68}_{8.6}$   & $48\>^{72}_{39}$  $^{95}_{17.2}$ & $100\>_{75}$  $_{35}$   & $100\>_{81}$  $_{44}$  \\

\hline

\vspace{0.5em}
$8-10$ (inc Kepler-34b) &  $3.1\>^{5.4}_{2.7}$  $^{9.3}_{1.1}$   &  $5.8\>^{9.9}_{4.9}$  $^{17.0}_{2.1}$   & $10.2\>^{16.8}_{8.4}$  $^{29}_{3.5}$   & $20\>^{34}_{17.1}$  $^{58}_{7.3}$    & $51\>^{73}_{39}$  $^{95}_{17.0}$   & $100\>_{78}$  $_{38}$  \\

\vspace{0.5em}

$6-10$ (inc Kepler-34b) &  $5.5\>^{8.3}_{4.5}$  $^{13.5}_{2.1}$   & $10.0\>^{15.5}_{8.3}$  $^{25}_{4.0}$   & $17.0\>^{27}_{14.3}$  $^{43}_{6.9}$   & $35\>^{54}_{29}$  $^{82}_{14.2}$   & $87\>^{86}_{56}$  $^{98}_{29}$   & $100\>_{83}$  $_{49}$ \\

\hline
\end{tabular}
\end{table*}

\begin{table*}
\caption{Percentage rates of occurrence for planets within 300\ d. Values are maximum likelihoods, with the rates of occurrence corresponding to 50 and 95\% confidence intervals shown as super and subscripts.}
\label{taboccrates300}
\begin{tabular}{@{}llllllr@{}}
\hline
R$_\textrm{planet}$   & \multicolumn{3}{|l|}{Planetary Inclination Distribution} \\
(R$_\oplus$) &  Coplanar  &  Gauss $\sigma=5^\circ$  & Gauss $\sigma=10^\circ$  & Gauss $\sigma=20^\circ$  & Gauss $\sigma=40^\circ$  & Isotropic \\
\hline
\hline
\vspace{0.5em}
$>10$ &  $0\>^{1.4}$  $^{5.9}$  &    $0\>^{3.4}$  $^{14.8}$  &  $0\>^{7.1}$  $^{31}$  & $0\>^{16.8}$  $^{67}$  & $0\>^{33}$  $^{89}$ & $0\>^{45}$  $^{93}$  \\

\vspace{0.5em}

$8-10$ &  $6.4\>^{11.6}_{5.7}$  $^{19.7}_{2.5}$   &   $17.5\>^{31}_{15.4}$  $^{53}_{6.6}$   & $38\>^{62}_{32}$  $^{91}_{13.7}$  & $100\>_{71}$  $_{32}$  & $100\>_{80}$  $_{42}$  & $100\>_{83}$  $_{44}$ \\

\vspace{0.5em}

$6-8$ &  $3.5\>^{8.5}_{3.1}$  $^{17.6}_{0.75}$  &  $10\>^{25}_{8.9}$  $^{51}_{2.2}$  & $21\>^{18.5}_{50}$  $^{88}_{4.7}$ & $47\>^{75}_{34}$  $^{96}_{9.1}$ & $100\>_{65}$  $_{18.4}$ & $100\>_{69}$  $_{21}$ \\

\vspace{0.5em}

$4-6$ &  $14\>^{26}_{11.7}$  $^{48}_{4.2}$  &  $51\>^{73}_{36}$  $^{96}_{13.4}$  & $100\>_{70}$  $_{27}$ & $100\>_{76}$  $_{32}$ & $100\>_{77}$  $_{34}$ & $100\>_{78}$  $_{35}$ \\

\vspace{0.5em}

$6-10$ &  $10.0\>^{16.3}_{8.2}$  $^{28}_{3.5}$  &  $30\>^{47}_{23}$  $^{77}_{10.0}$  & $58\>^{78}_{44}$  $^{96}_{19.7}$  & $100\>_{77}$  $_{37}$ & $100\>_{81}$  $_{43}$  & $100\>_{83}$  $_{47}$  \\

\vspace{0.5em}

$4-10$ &  $20\>^{34}_{17.2}$  $^{58}_{7.4}$  &  $67\>^{82}_{49}$  $^{98}_{22}$  & $100\>_{78}$  $_{38}$ & $100\>_{81}$  $_{44}$ & $100\>_{82}$  $_{44}$ & $100\>_{85}$  $_{50}$ \\

\hline
\end{tabular}
\end{table*}

\begin{figure}

\resizebox{\hsize}{!}{\includegraphics{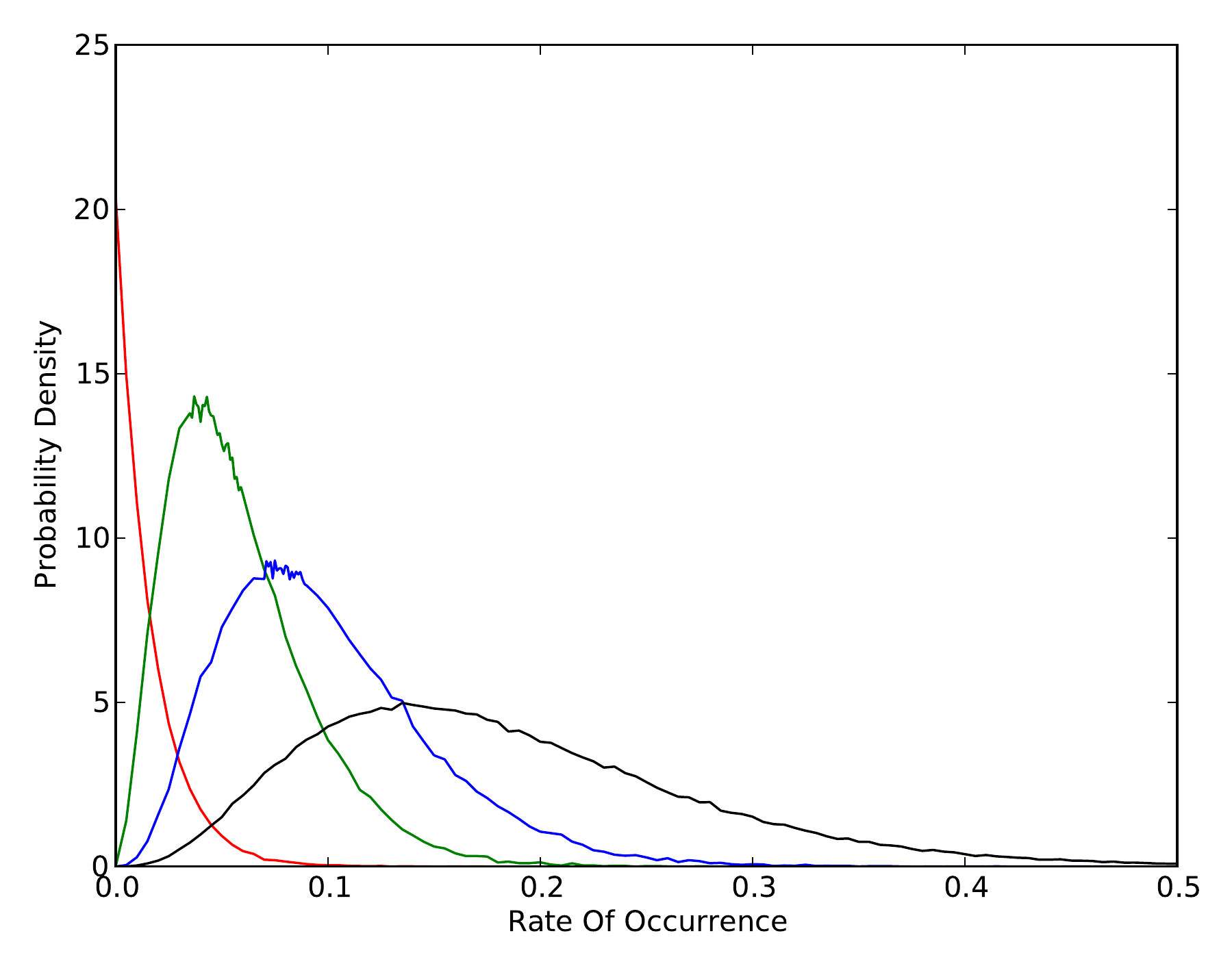}}
\caption{Probability density functions for the rate of occurrence of CB planets following a Gaussian inclination distribution with $\sigma=5^\circ$, within 10.2$P_\textrm{bin}$. The distributions are shown for (from left to right) planets with radii ${>}10R_\oplus$, 8-10$R_\oplus$, 6-10$R_\oplus$, and 4-10$R_\oplus$. The ${>}10R_\oplus$ density function has been scaled down by a factor of three for clarity, and takes a different form to the others due to the zero detections of planets within this group.}
\label{figdists}
\end{figure}

Using these detections and the debiased sample of Section \ref{SectDebiasing}, we can obtain probability density functions of the implied circumbinary planet rate of occurrence in the Kepler sample, around non-contact binary stars. Typical such distributions are shown in Figure \ref{figdists}. These are non-Gaussian, and so we present values along with 50 and 95\% confidence limits. The specific values and errors were found to be only moderately sensitive to the presence of a pile up in planet periods near the inner stability limit. Without full confirmation of its existence we choose to present values without this pileup, but including one (through recovering 50\% of unstable planets into the pileup region as described in Section \ref{sectpopsynthplanetdists}) leads to occurrence rates which are \mytilde10\% lower for the 300\ d period, coplanar group, and unchanged for the 10.2 P$_\textrm{bin}$ group. These further reduce in significance for more uniform inclination distributions, and are well within the 50\% confidence limits. 

\begin{figure}

\resizebox{\hsize}{!}{\includegraphics{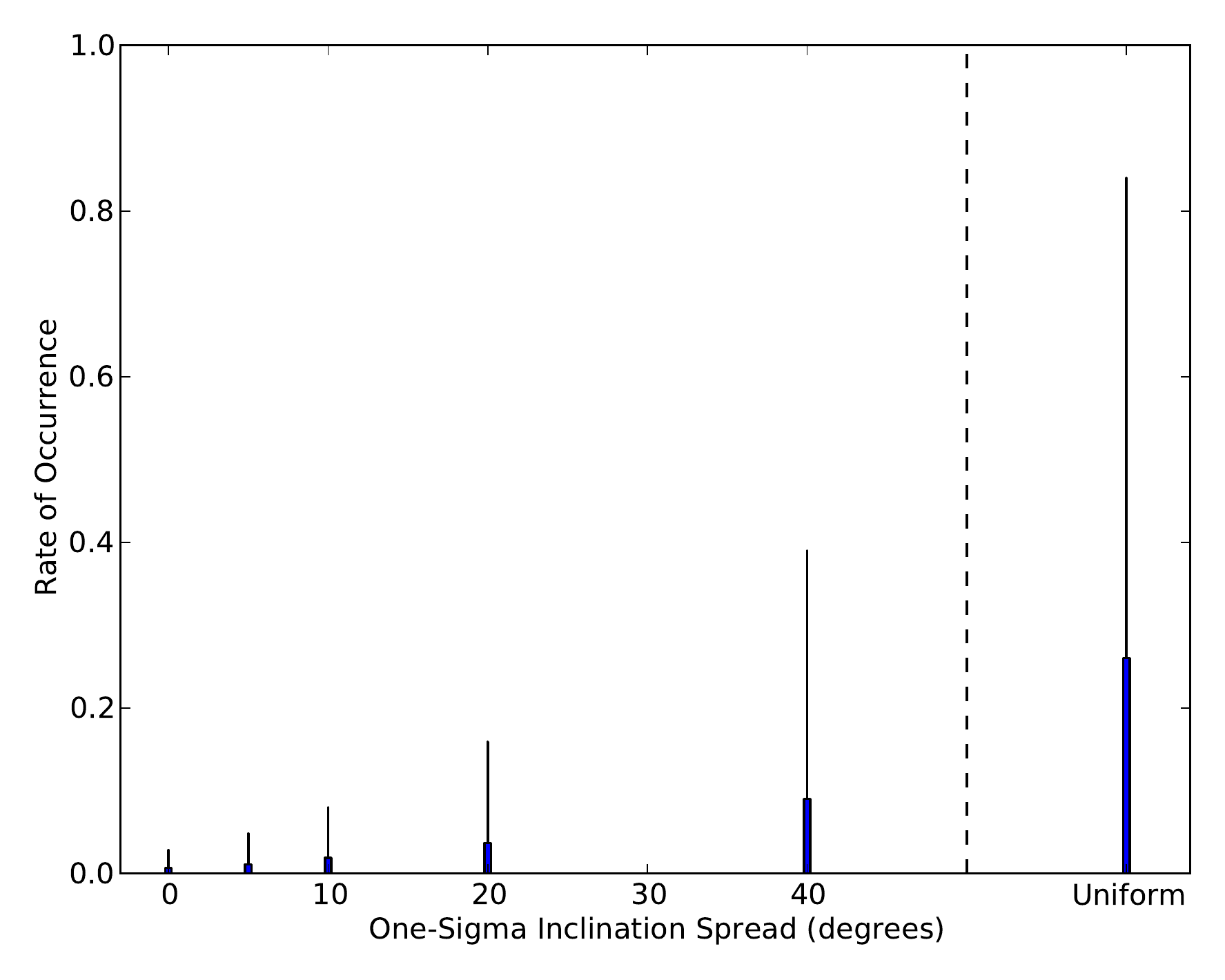}}
\caption{Rates of occurrence for a range of Gaussian planetary inclination distributions, for planets within 10.2$P_\textrm{bin}$ with $R_p>10R_\oplus$. The large boxes show 50\% confidence limits, with the thin `whiskers' extending to 95\% limits.}
\label{figAb10}
\end{figure}

\begin{figure}

\resizebox{\hsize}{!}{\includegraphics{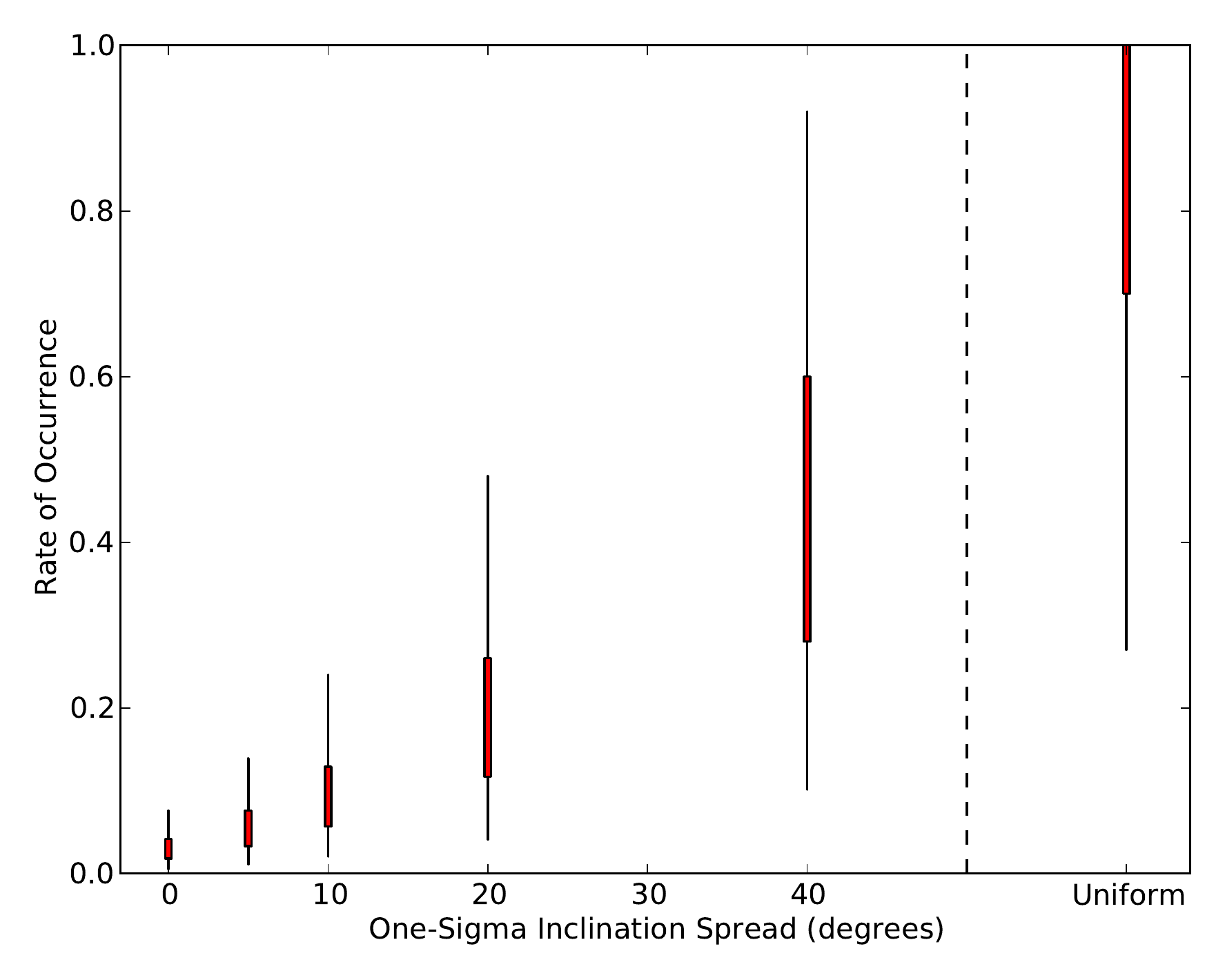}}
\caption{As Figure \ref{figAb10} for $8<R_p<10R_\oplus$}
\label{figAb8}
\end{figure}

\begin{figure}

\resizebox{\hsize}{!}{\includegraphics{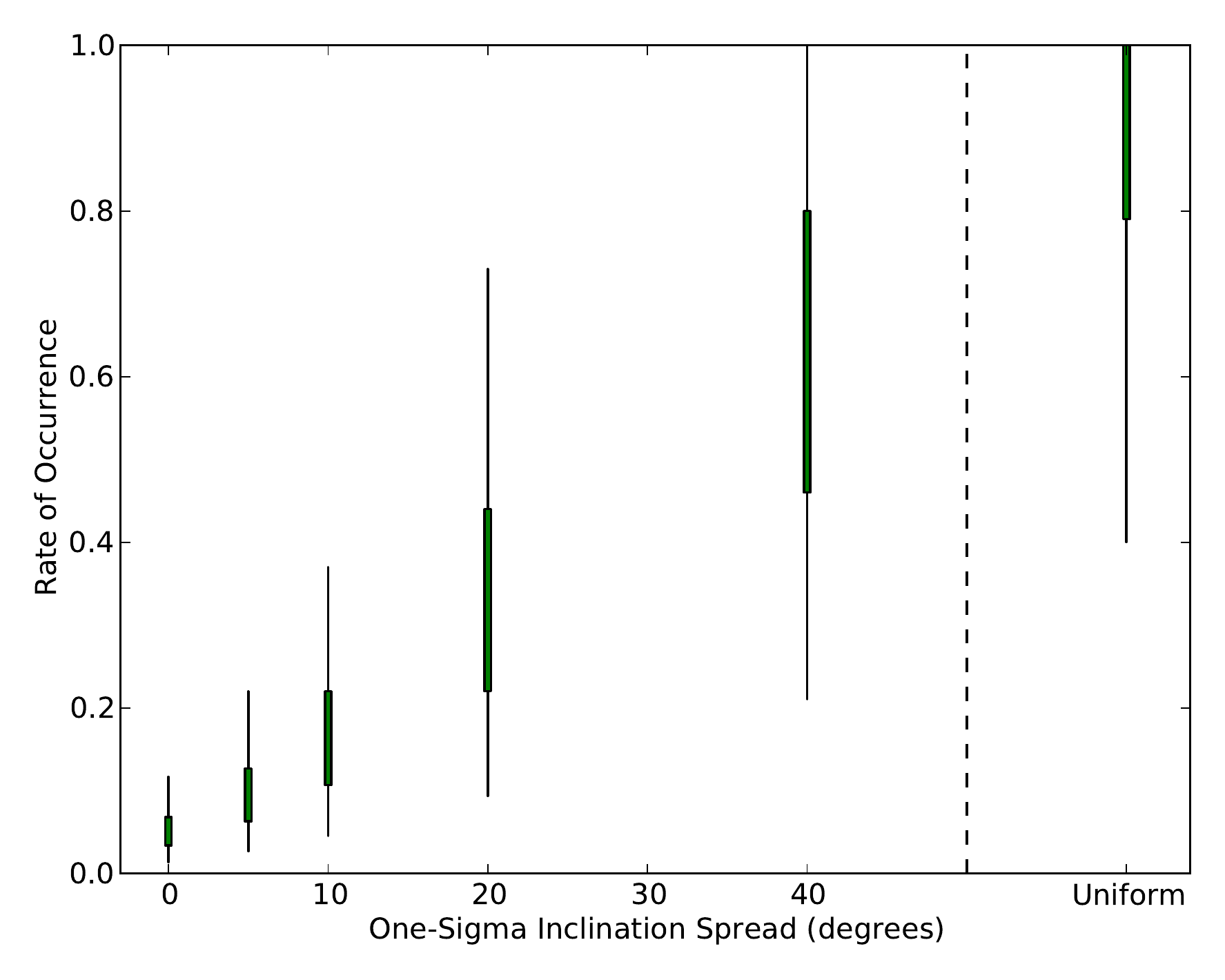}}
\caption{As Figure \ref{figAb10} for $6<R_p<10R_\oplus$}
\label{figAb6}
\end{figure}

\begin{figure}

\resizebox{\hsize}{!}{\includegraphics{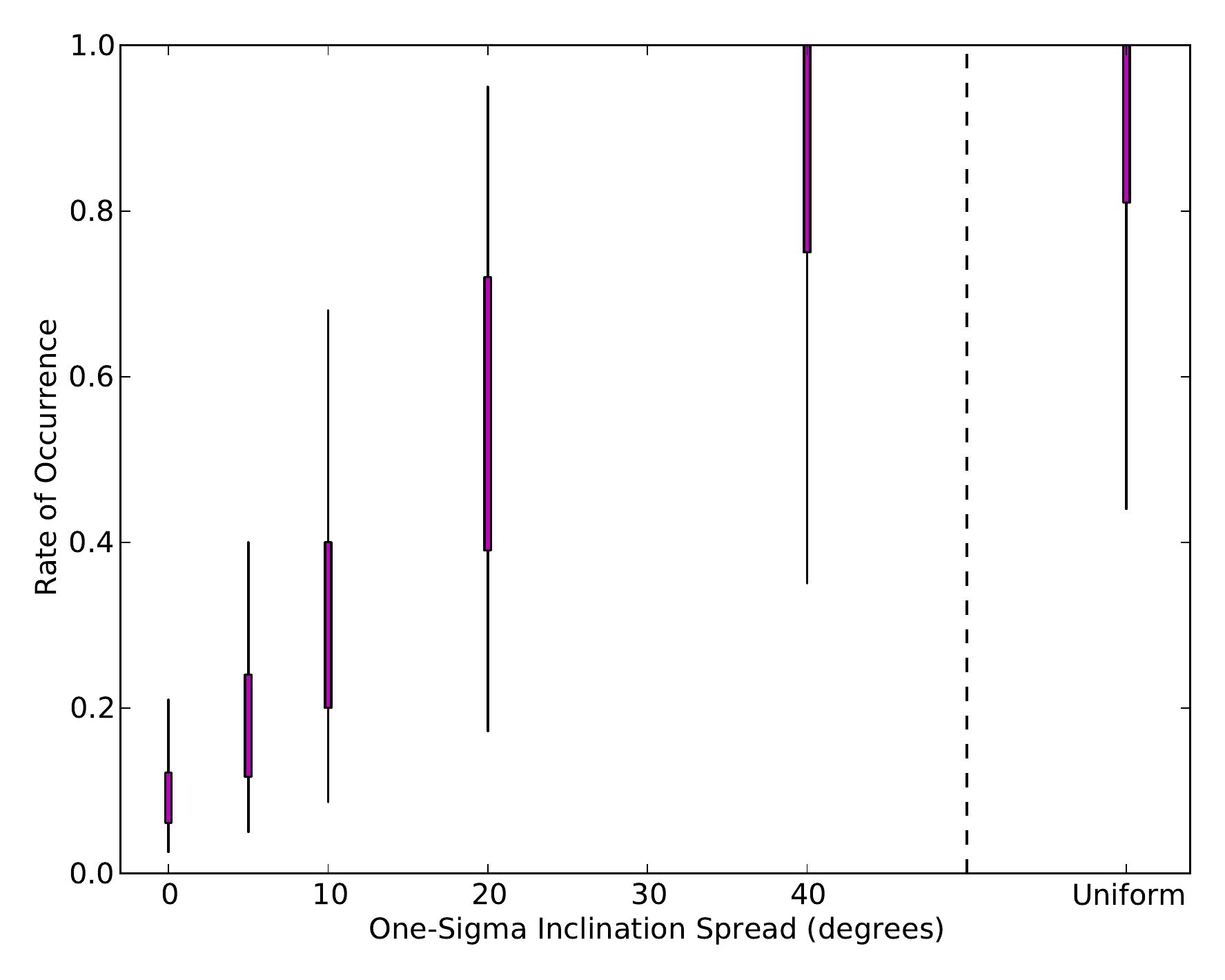}}
\caption{As Figure \ref{figAb10} for $4<R_p<10R_\oplus$}
\label{figAb4}
\end{figure}

The occurrence rates are however critically dependent on the input planetary inclination distribution. As such results are shown as a function of this, and are summarised in Figures \ref{figAb10}--\ref{figAb4}. The full list of values and confidence limits can be seen in Tables \ref{taboccrates102} and \ref{taboccrates300} (note that modal values are typically accurate to \mytilde 0.5\%, unless higher precision is given). The rates in Table \ref{taboccrates102} are lower (and more precise) than for Table \ref{taboccrates300} as 10.2 P$_\textrm{bin}$ is generally lower than 300\ d in the Kepler sample. This improves transit detection, increasing the sample size of binaries while not increasing the planet count, as nearly all planets are still detectable at 300\ d and Kepler-34b no longer lies within the period window. This concentration of the known transiting circumbinary planets at periods close to the binary has been discussed in Section \ref{sectpopsynthplanetdists}. The varying rates are then a consequence of the window on parameter space one uses to look at the sample.

We show a number of planet radius bins, both large and small, so that readers may use whichever is most useful for their science. For the periods below 10.2 P$_\textrm{bin}$ we present results both with and without Kepler-34b. Strictly Kepler-34b lies at 10.4$P_\textrm{bin}$, just above the period threshold. In the case of CBs it seems plausible however that a more suitable boundary would be defined by multiples of the binary inner stability limit. In the Kepler-34 case, this limit is particularly large, at \mytilde190\ d, due to the high eccentricity of the binary. Under this definition, Kepler-34b would clearly lie within a similarly defined period boundary. As such we present both results where relevant.

Although a large range of planetary inclination distributions is tested, previous work suggests that some are more likely than others, and that a strong preference for coplanarity is probable \citep{Foucart:2013gk}. Using the coplanar results as an indicative case, we find that there is a 95\% confidence upper limit on the occurrence rate of giant (${>}10R_\oplus$) planets within 10.2 $P_\textrm{bin}$ of 2.8\%. Making comparisons to the single star rate of occurrence \citep{Fressin:2013df} is difficult, as we do not use the same period ranges. However, looking at their largest two ranges, 0.8--245\ d and 0.8--418\ d, these rates are $\mytilde 5$\% for planets with $R_p>6R_\oplus$ and $\mytilde 8$\% for planets with $R_p>4R_\oplus$, the latter derived by summing the appropriate radius bins in their paper. Both of these are consistent with our coplanar results, although our modal values are higher. It is worth noting that were we to assume the single star rate of occurrence holds in the circumbinary case, for the $>6R_\oplus$, within 300\ d bin the 10$^\circ$ Gaussian inclination distribution would be excluded with probability ${>}99.9$\%, along with all more misaligned distributions. As such, should a large very misaligned population of circumbinary planets exist, it would imply that circumbinary planets exist in significantly greater numbers than planets with single stellar hosts.

The derived probability density functions also allow us to investigate differences between planetary radius groups. It has been proposed that giant (Jupiter like) planets should be less common in coplanar circumbinary orbits than Saturn-like or smaller equivalents, due to increased chances of ejection for higher mass planets \citep{Pierens:2008fq}. We find that the Kepler sample supports this, with the rate of occurrence for planets ${>}10R_\oplus$ within 300\ d being significantly lower than the other radius groups. In the coplanar case the significance of this difference is 99.8\% ($4-10R_\oplus$), 98.4\% ($6-10R_\oplus$), and 96.4\% ($8-10R_\oplus$) (these bins are used for comparison because of their higher planet count). This finding becomes less significant for distributions more misaligned than the 10$^\circ$ Gaussian case.

Finally, it has also been proposed that there is a preference for CBs to have longer period binary hosts \citep{Welsh:2013wo}. All of the known planets so far orbit binaries with periods greater than 7d, despite these longer period binaries being significantly undersampled in the Kepler dataset. We are able to test whether this effect is due to a sampling bias or represents a real trend using our debiased sample. We split the sample into short and long period binaries, using a period cut of 10\ d. For coplanar CB planets with periods less than 10.2$P_\textrm{bin}$, we find the probability that the occurrence rate is lower around shorter period binaries to be 96.3\%(4-10$R_\oplus$), 97.7\%(6-10$R_\oplus$) and 95.6\%(8-10$R_\oplus$). This becomes more significant for more misaligned inclination distributions, rising to 99.9\% for the 5$^\circ$ Gaussian case and higher. Using a binary period cutoff at seven days (below all published CB planets) reduces the significance of the result, to a 92.6\% probability for the 6-10$R_\oplus$ sample. This again becomes more significant for more misaligned distributions.

\subsection{Highly Inclined Distributions, and Multiple Planets/System}
As said earlier, we have used rates of occurrence to mean here the number of binaries with one or more planets as a proportion of the total binary number. This leads naturally to a maximum occurrence rate of 100\%. However, as can be seen in Tables \ref{taboccrates102} and \ref{taboccrates300}, some tested cases run into this limit, particularly the highly inclined planetary inclination distributions. This has some effect on the results. A large potential area under the probability density function curve in these cases can be found above 100\% (i.e. representing multiple planets per binary) and is excluded from our values and analysis due to this definition of the occurrence rate. While we have no wish to include multiple planets formally at this time (noting the additional search algorithm, planet parameter correlations, and dynamical questions which would need to be answered), it is informative to investigate the effects of these unused areas of the probability curves.

We have noted previously the particularly high occurrence rates required by highly inclined distributions such as the isotropic case. Allowing multiple planets per binary, the full extent of this issue can be demonstrated. We tested this by allowing the occurrence rate to rise above unity in our model (keeping all planet parameters independent). In a typical high inclination case ($P < 10.2P_{\textrm{binary}}$, 8-10$R_\oplus$, Isotropic, without Kepler-34b) the results rise to $113^{418}_{36}$\% (with the values corresponding to 95\% confidence limits super and subscripted), showing a strong preference for more than one planet per system. In the most extreme case ($P < 300$\ d, 4-10$R_\oplus$, Isotropic) the results rise dramatically to a modal value of near 50 planets per binary, a number which would presumably lead to serious stability issues within this relatively tight period bound. Note that due to the change in definition of occurrence rate implied here these numbers cannot be considered a direct extension of the previous results, and are merely indicative. When values are needed, those given in Tables \ref{taboccrates102} and \ref{taboccrates300} should be used with the earlier definition of the occurrence rate. In the light of this however, we repeat that should the true inclination distribution of circumbinary planets be particularly misaligned with respect to their host binaries, their formation must be abundant, common and in essence very hard to avoid.

\begin{figure}
\resizebox{\hsize}{!}{\includegraphics[trim=0.6cm 0.5cm 0.5cm 0cm]{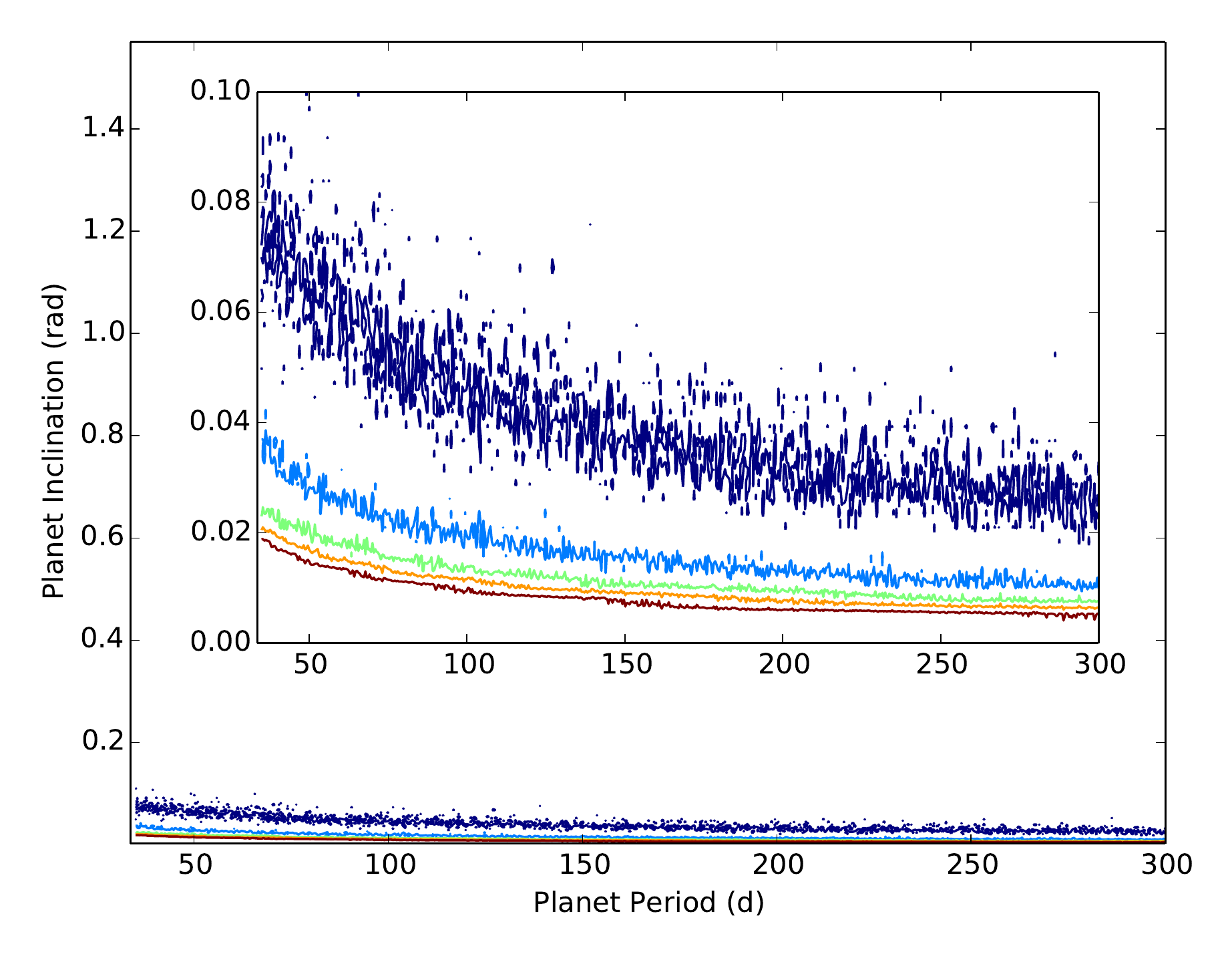}}
\caption{Contours of the proportion of planets showing consecutive transits, drawn from a uniform sample of planets with the shown range of inclinations and periods orbiting a binary (itself with inclination $\pi$/2, period 8.5d, and solar radii and mass stars). Contours are plotted at 20\% intervals, starting at 10\%. for the uppermost line.}
\label{figinceffect}
\end{figure}

\section{Discussion}

\subsection{Rate of Occurrence and Errors}
The errors associated with our presented rates of occurrence are particularly large compared with those for single stars; this is a function of both the reduced sample size and moreover the constraint of consecutive transits. The region of parameter space within which consecutive transits occur is decidedly smaller than for single stars, reducing the sensitivity of a given sample. Both these occurrence rates and their errors increase sharply for increasingly uniform planetary inclination distributions. This behaviour is expected, as in the uniform case many more planets must exist in order to produce the few we see transiting. The errors increase as the possible parameter space of planetary orbits becomes largely unprobed by our consecutive transit requirement, which is only sensitive to nearly coplanar planets. To illustrate how this situation comes about, we plot our sensitivity as a function of planetary inclination in Figure \ref{figinceffect}. The region of parameter space where consecutive transits are possible is shown. The tiny area of the total parameter space this represents is striking - that a reasonable number of planets should be found within it (as they have been) largely explains the qualitative form of our results. Unbiased searches for misaligned CB planets, for example on non-eclipsing binary stars \citep{Martin:2014ug}, will be essential to constraining the CB inclination distribution and the implied rates of occurrence. Interestingly, the occurrence rate estimated in that work is compatible with our values, despite being based on a different method and involving no analysis of the Kepler light curves themselves.

We note that the presented values test planets after significant periods of evolution. It is unlikely that any of the known transiting CB planets formed where they are currently located \citep{Meschiari:2012er}. As such these rates include both planet formation and subsequent dynamical evolution, through disk migration, scattering or otherwise. Furthermore the starting point of this history is not fully understood - the abundance of circumbinary disks is not yet well known, although it has been shown that they should be common \citep{Alexander:2012cn}. If these disks occur more or less readily than circumstellar disks then it impacts the formation rates implied by our presented rates of occurrence.

\subsection{Biases and Approximations}
Any statistical study is subject to various potential biases, which we summarise here. The first is in the sample chosen, of Kepler eclipsing binary stars. This is not a general sample of binaries, with a study of the full effects of the Kepler pipeline well beyond this paper. We are also biased towards shorter period binaries, the usual geometric bias associated with selecting eclipsing objects. As such our rates of occurrence are skewed towards these shorter period binaries. Given that the currently known transiting CBs are found orbiting generally longer period binaries ($P_\textrm{bin}\gtrsim 5$d) this may be significant, and the effects of this will be the target of future work. 

There is also a bias against more active stars (with noisier light curves) due to the difficulty in detecting planetary transits, especially where the timescale of that noise becomes shorter than \mytilde3\ d (see Section \ref{sectindieventsearch}). This will preferentially reject closer binaries, as they are more likely to have activity induced by the companion, and so leads to a sample bias towards longer period binaries within the dataset. Stars with particularly sharp binary eclipses may also be affected, although the effect of these eclipses has been mitigated as far as possible (see Section \ref{sectebsigremoval}). Similarly, planets with orbits on very close integer resonances with the host binary are more likely to be rejected as noise, or to have their transits removed with the binary signal. 

In checking for consecutive transits, a Keplerian approximation was made as to the planetary orbits. This will become important for planets orbiting on short enough periods that their precession timescales become comparable to the data baseline (\mytilde4\ yr). Using the formula of \citet{Doolin:2011ib}, derived from \citet{Farago:2010ev}, we can determine where this region typically begins: for a moderately eccentric $e_p=0.2$ coplanar planet at the inner stability limit, orbiting an $e_\textrm{bin}=0.1$ binary, the binary must have a period under \mytilde0.06\ d for the planetary precession period to fall below 4 years. As such, this will not be a problem here. The precession timescales of CB planets are however fairly short, on the order of decades \citep{Armstrong:2013kg}. This means that objects which consecutively transit through the dataset may well not do in several years time, as is the case for Kepler-16b \citep{Doyle:2011ev}. This is accounted for by the statistical nature of our method -- a planet on a slightly misaligned orbit will consecutively transit for a fraction of the iterations, and only be counted for that fraction. As detailed in the above section, the consecutive transit requirement also impacts our sensitivity to inclined planets. This is included in our presented errors, but shows that the information leading to our results comes from a narrow region of parameter space in terms of planetary inclination.

We have also not accounted for tertiary stellar companions, which will dilute planetary transits and reduce the chance of detection. This contamination has been estimated to be potentially as high as 20\% \citep{Rappaport:2013de}. Without further detail it is impossible to estimate how strongly such tertiary companions would dilute transits of the primary star, so we prefer to produce rates of occurrence without this. In the worst case (if in 20\% of our sample the dilution was always strong enough that we could not in fact detect planets) the true sample size would be reduced by this 20\%. This would have the effect of increasing the presented occurrence rates by \mytilde 20\% of their present values. This would not affect the conclusions made above.

Similarly in terms of the injected transits, transits of the secondary star were not included. We do not expect these to contribute significantly to the detection. Dilution by the primary star means that the relative depth of transits of the secondary star compared to those on the primary goes as $(T_2/T_1)^4$, implying that for all but particularly equal temperature binaries ($T_2/T_1>\mytilde 0.92$, corresponding to a transit depth ratio of \mytilde 0.7) transits of the secondary would not contribute significantly to the detection. From \citet{Raghavan:2010gd}, their figure 16, it is possible to estimate how many binaries this applies to. This estimate is somewhat rough (as the samples are by no means the same, and it involves converting mass ratio to temperature) but leads to \mytilde 10-15\% of the sample having significant secondaries. As several of these binaries will already be successful detections, including secondaries would increase the detectable binary sample by at most a few percent, decreasing the derived abundance rates by a few percent of their present values.

We have relied on an element of human eyeballing of the search algorithm results, introducing potential subjectivity. This was mitigated through using two independent checkers, and the results supported in that every known planet host (excepting Kepler-47, which the algorithm did not detect) was marked by both. The use of defined significance thresholds (see Section \ref{SectTInjections}) also constrained the sample to a size amenable to finely detailed checking.

Finally, there is a possible effect from errors on the temperatures of \citet{Armstrong:2013cp}; these are \mytilde 400K for the primary stars and \mytilde 600K for the secondaries, which would affect the derived radii used to produce transit depths and check for consecutive transits. As our results are statistical, errors on individual binaries will not have a large effect, the important factor being whether systematic offsets are found in the temperatures. We cannot check for this, but there is no indication that they should be present.

\section{Summary and Conclusion}
We have investigated the rates of occurrence of circumbinary planets orbiting close ($P_\textrm{bin}<$\mytilde60\ d) non-contact binary stars using the Kepler sample of eclipsing binaries. This produced a number of interesting results:

\begin{enumerate}
\item The most significant controlling distribution is that of planetary inclination - whether these planets lie preferentially coplanar with their host binaries, or in a more uniform pattern. Our results show that if such a uniform or even generally misaligned distribution is the norm, then the rate of occurrence of CBs must be exceptionally high, significantly more so than analogous rates for single stars. While not formally excluding very uniform, misaligned planetary inclination distributions, these results show that to exist such distributions need planetary formation rates at levels very difficult to explain. 

\item Conversely, if coplanarity is preferred, to the level implied by a Gaussian distribution with standard deviation \mytilde $5^\circ$ or tighter (although we note that the distribution by no means must be gaussian, and may even be bimodal) then the rate of occurrence of CBs is consistent with that of single star planets. Evidence suggests that circumbinary planets orbiting sub-AU binaries should be preferentially coplanar due to alignment of the protoplanetary disk, supporting this option \citep{Foucart:2013gk,Kennedy:2012jl}.

\item CB giant planets (defined as $>=10R_\oplus$) are significantly less common than their smaller equivalents. There remains the possibility of a non-coplanar giant CB population at any rate of occurrence, formed for example by dynamical evolution, but a coplanar CB giant population on the same order as planets with $R<10R_\oplus$ is excluded, at least within our tested period range. Given that proto-planetary disk masses scale with the mass of the central object \citep{Andrews:2013ku}, and that more massive disks produce more gas giants \citep{Mordasini:2012ek} this supports the finding of \citet{Pierens:2008fq}, that CB Jupiter mass planets if present will likely orbit at larger distances from the central binary due to increased scattering.

\item CB planets are less common in coplanar orbits around shorter period binaries ($P_\textrm{bin}<\mytilde\> 5-10$\ d) than around binaries of longer period. We have shown that this trend is not the result of sampling bias, with 99.9\% confidence for all tested misaligned planetary inclination distributions and 97.7\% for a coplanar distribution. The observed difference could be explained through a significantly different orbital distribution between planets orbiting shorter and longer period binaries (such as a more misaligned shorter population, so that we do not observe them) or by an effect of the formation of these binary systems \citep[see e.g.][]{Fabrycky:2007jh}. If shorter period binaries form through secular interactions with a tertiary stellar companion, planets in these systems would either be disrupted, or if present difficult to see due to dilution by the companion. If such close binaries have evolved to their current orbit via angular moment loss (through e.g. magnetic braking) then this process may influence the protoplanetary disk and impact planet formation. This remains a promising area of future work.

\end{enumerate}

To improve our knowledge of these unusual systems a larger sample of circumbinary planets needs to be found. Fortunately there are several possible routes to these discoveries, from searches for misaligned transiting planets to the use of radial velocities or binary eclipse timing. All of these will help to increase the sample size available, leading to new insights into their formation, evolution, and how these impact on general planet formation theories. The discovery of more misaligned planets will allow tighter constraints to be placed on planetary inclination distributions, answering questions about the dynamical evolution of these systems. Beyond this, future space missions such as PLATO and TESS should provide a great deal more new transiting, bright CB planets for further work.

\section*{Acknowledgements}
We would like to thank the reviewer, Darin Ragozzine, for thorough and helpful comments on the manuscript. DJA would like to thank Thomas Marsh for discussions which greatly aided this work. This paper includes data collected by the Kepler mission. Funding for the Kepler mission is provided by the NASA Science Mission Directorate.

\bibliography{papers020414}
\bibliographystyle{mn2e_fix}

\end{document}